# Broadband dielectric spectroscopy and aging of glass formers

R. Wehn, P. Lunkenheimer[*], A. Loidl

*Experimental Physics V, Center for Electronic Correlations and Magnetism, University of Augsburg, 86135 Augsburg, Germany*

**Abstract**

We present the results of aging experiments on a variety of glass formers, including xylitol, glycerol, propylene carbonate, and $[Ca(NO_3)_2]_{0.4}[KNO_3]_{0.6}$ (CKN). In addition, broadband dielectric spectra of xylitol and CKN are provided. We demonstrate that, irrespective of which dielectric quantities (dielectric constant, loss and modulus) are analyzed, and irrespective of the spectral region where they have been measured, the aging-time dependence always is governed by the structural rearrangements constituting the α-relaxation. If the time dependence of the structural relaxation time during aging is taken into account, relaxation times and stretching parameters, fully consistent with equilibrium data, are obtained. In CKN, both the structural relaxation time and the strongly decoupled conductivity relaxation time are deduced from the same dielectric aging experiment.

## 1. Introduction

In addition to the hallmark features of glassy matter, stretching, non-Arrhenius behavior, and ergodicity breaking [1], in recent years also a variety of further unusual phenomena considered as common to glassy materials were discovered, mostly by using broadband dielectric spectroscopy [2]. Among them are, e.g., the occurrence of secondary relaxations [3], which sometimes lead to a so-called excess wing in the spectra of the loss [2,4,5,6], and an additional high-frequency contribution in the GHz-THz region [2,7,8]. In addition to the typical dielectric experiment leading to information on the response of glassy matter to electrical excitations (e.g., an ac field), also the response to a temperature variation (e.g., a downward or upward temperature step), can be conveniently investigated by dielectric methods (see, e.g. [9,10,11,12,13]). In the present work, amongst others, we show results from applying both approaches on xylitol, a typical glass former having a well-defined β-relaxation and on $([Ca(NO_3)_2]_{0.4}[KNO_3]_{0.6}$ (CKN) an ionic-melt glass former.

Special emphasis is put on the so-called "physical aging" [14], which here is investigated by monitoring the time-dependent variation of various dielectric quantities after the sample was quickly cooled to a temperature below the glass temperature $T_g$. In two recent works [12,15], we demonstrated that the time dependence of the imaginary parts of the dielectric loss ε" (for glass formers comprising dipolar molecules) and the dielectric modulus M" (for glass formers dominated by ionic conductivity) can be described by the same parameters as the structural relaxation process, irrespective of the dynamic process prevailing in the investigated frequency region. In particular, the same relaxation time and non-exponentiality parameter as resulting from equilibrium measurements is obtained, if only the time dependence of the relaxation time during aging is properly taken into account. In the present work, we provide additional data also on the aging-time dependence of the dielectric constant ε' of several glass formers. In addition, the aging of CKN is analyzed in both, the dielectric loss and modulus representation. To better understand the aging behavior in xylitol and CKN, also broadband spectra of these two materials are provided. In all cases, a common time dependence of all quantities and all processes for a single material is revealed. The relaxation times obtained from the aging experiments are compared to those from the equilibrium experiments.

---

[*] Corresponding author. Tel. +49 821 598 3603; fax: +49 821 598 3649; E-mail: peter.lunkenheimer@physik.uni-augsburg.de

## 2. Experimental procedures

For the low-frequency and the aging measurements, parallel plane capacitors having an empty capacitance up to 100 pF were used. High-precision measurements of the dielectric permittivity in the frequency range $10^{-4} \leq \nu \leq 10^{6}$ Hz were performed using a frequency response analyzer. At selected temperatures and aging times, additional frequency sweeps at $20$ Hz $\leq \nu \leq 10^{6}$ Hz were performed with an autobalance bridge [16]. To keep the samples at a fixed temperature for up to ten weeks, a closed-cycle refrigerator system was used for temperatures below ambient and a self-made oven for the higher temperatures. The samples were cooled from a temperature at least 20 K above $T_g$ with a cooling rate $\geq 3$ K/min. The final temperature was reached without any temperature undershoot. As zero point of the aging times $t_{age}$, we took the time when the desired temperature was reached, typically about 100 s after passing $T_g$. The temperature was kept stable better than 0.1 K for all aging measurements. For details on the experimental setups used at high frequencies, see [16].

## 3. Results

Figure 1 shows the aging-time dependence of both, the dielectric constant and the dielectric loss of three typical molecular glass formers after quenching them to a temperature several K below the glass temperature. Glycerol ($T_g \approx 185$ K) and propylene carbonate (PC, $T_g \approx 159$ K) belong to the group of glass formers (termed "type A" [17] or "EW" [18]) that show a second power-law decrease on the high-frequency flank of the α-relaxation loss peak [2,4,5,6,17,19,20]. This phenomenon was termed "excess wing" in [21] to account for the intensity in excess of the high-frequency flank of the α-relaxation peak, connected with this spectral feature. In contrast, xylitol ($T_g \approx 248$ K) is known to exhibit a well-pronounced secondary relaxation ("type B" glass former [17]) [22,23,24,25,26]. From the results reported in [26], by using a criterion [18] for genuine Johari-Goldstein (JG) relaxations [3] based on Kia Ngai's coupling model [27], it can be concluded that the β-relaxation in xylitol indeed is of JG type. The occurrence of JG relaxations is commonly assumed to be a property inherent to the glassy state of matter [1,3]. For all frequencies investigated, both, ε"(t) [12,15] and ε'(t) reveal a monotonous decrease during aging. The saturation reached at the longest aging times demonstrates that thermodynamic equilibrium is reached. In contrast to glycerol and PC, in the time-dependent loss of xylitol [Fig. 1(f)] there is a crossing of two curves, namely of those at 1 kHz and 10 Hz.

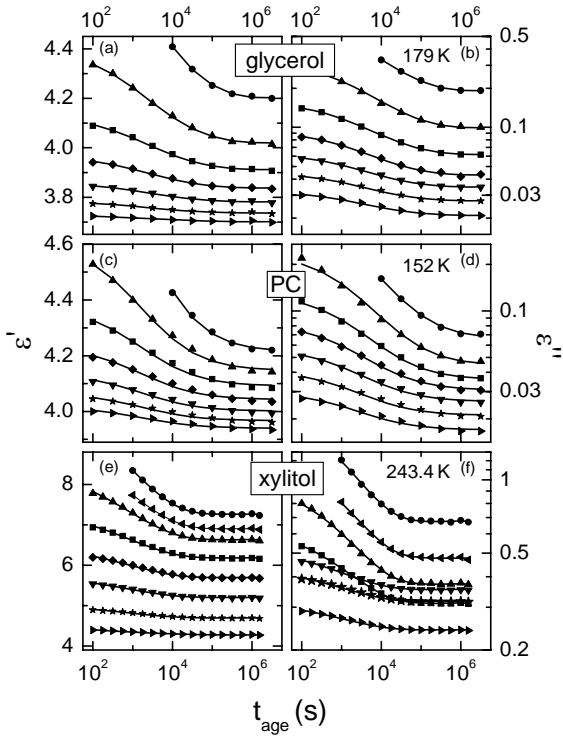

Fig. 1. Aging-time dependence of the dielectric constant (a,c,e) and loss (b,d,f) of glycerol (a,b), propylene carbonate (c,d), and xylitol (e,f). The data were collected at the temperatures indicated in the figures and at the following measurement frequencies (for better readability, not all frequencies are shown in all frames): 0.1 Hz (circles), 0.3 Hz (triangles pointing left), 1 Hz (triangles up), 10 Hz (squares), 100 Hz (lozenges), 1 kHz (triangles down), 10 kHz (stars), and 100 kHz (triangles right). The lines are fits using the modified KWW law including a time-dependent relaxation time, as promoted in [12] [eqs. (1) and (2)]. For each material, all curves ($\varepsilon'$ and $\varepsilon''$ and all frequencies) were fitted with identical $\beta_{age}$ and $\tau_{age}(t)$. The experimental errors are of similar magnitude as the width of the symbols.

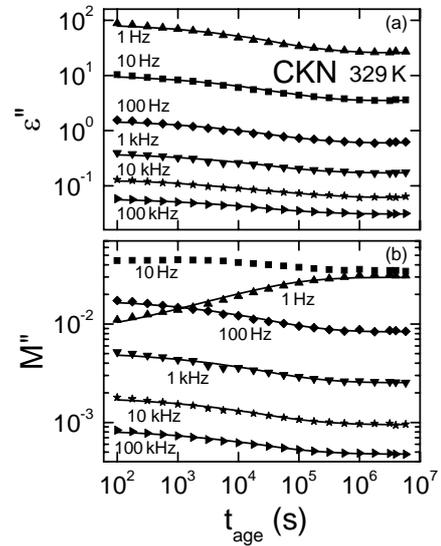

Fig. 2. Aging-time dependence of the dielectric loss (a) and modulus (b) of CKN measured at 329 K and various frequencies. The lines are fits using the modified KWW law including a time-dependent relaxation time, as promoted in [12] [eqs. (1) and (2)]. For each material, all curves ($\varepsilon'$ and $\varepsilon''$ and all frequencies) were fitted with identical $\beta_{age}$ and $\tau_{age}(t)$. The experimental errors are of similar magnitude as the width of the symbols.

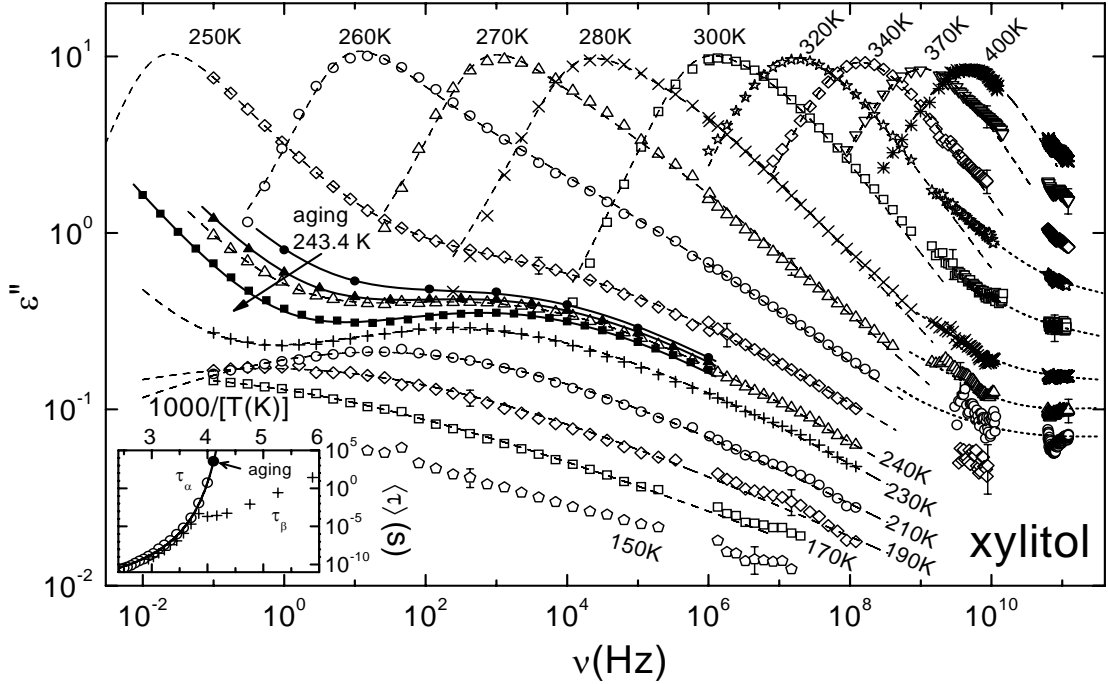

Fig. 3. Loss spectra of xylitol for various temperatures (open symbols). The sub-$T_g$ spectra at $T < T_g \approx$ 248 K were not taken in equilibrium. The solid symbols show three spectra collected during the aging experiment performed at 243.4 K at aging times of $10^2$ s, $10^3$ s, and $10^{6.2}$ s (from top to bottom). The experimental errors are of similar magnitude as the width of the symbols, except for low temperature and high frequencies, where typical error bars are indicated. The solid and dashed lines are fits with the sum of the CD and CC function at $T \leq 300$ K and the HN and the CC function at $T \geq 320$ K. The dotted lines, drawn to guide the eye, indicate the likely presence of a loss minimum at high frequencies. The inset shows the average relaxation times [43] obtained from the fits of the spectra shown as open symbols in the main frame (circles). The closed circle in the inset shows the equilibrium structural relaxation time $\tau_{eq}$ obtained from the analysis of the aging experiment. The line is a fit of $\tau_{diel}(T)$ with a VFT law ($\tau_0 = 6.4 \times 10^{-14}$ s, B = 1290 K, and $T_{VF} = 210$ K). The pluses show the β-relaxation time.

In Fig. 2 the aging-time dependence of CKN is shown. In addition to the time-dependent dielectric modulus [12], now we also provide the dielectric loss. While ε'(t) decreases monotonously during aging, M"(t) exhibits the indication of a maximum at 10 Hz and even increases at 1 Hz. In all cases a saturation value is approached for the longest aging times when equilibrium is reached.

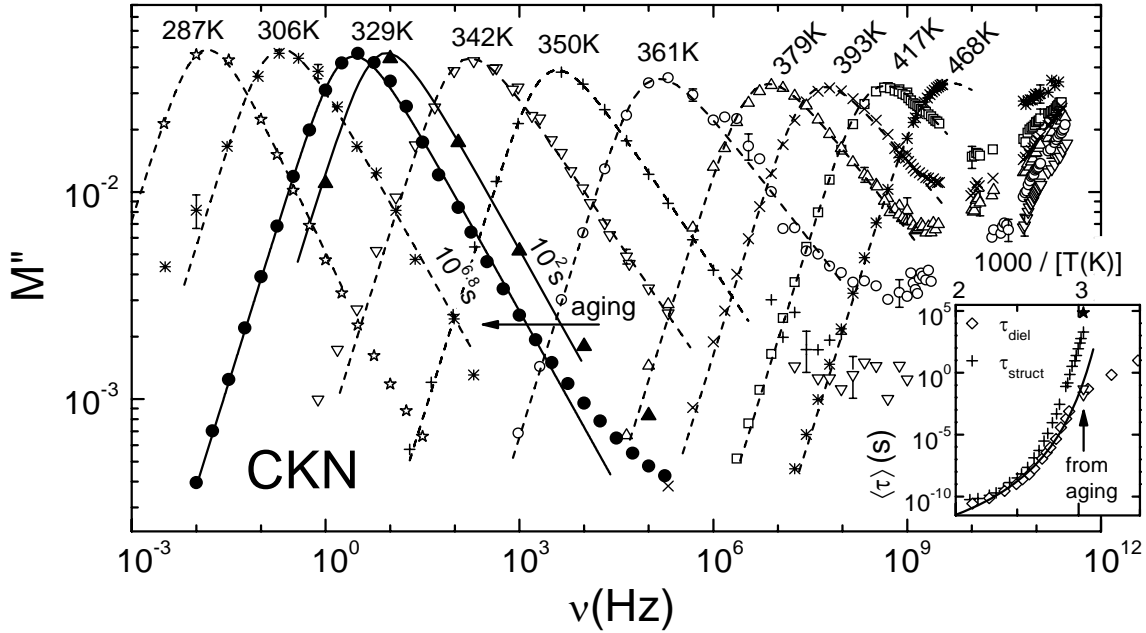

Fig. 4. Spectra of the imaginary part of the dielectric modulus of CKN for various temperatures (open symbols). The sub-$T_g$ spectra at 306 and 287 K were not taken in equilibrium. The solid symbols show the results of the first and last sweep collected during the aging experiments performed at 329 K. The experimental errors are of similar magnitude as the width of the symbols, except for the regions where typical error bars are indicated. The lines are fits with the CD function. The inset shows the average relaxation times $\tau_{diel}$ obtained from the fits of the spectra shown as open symbols in the main frame (lozenges) and of the two aging spectra (triangles). The upper triangle corresponds to the equilibrium dielectric relaxation time at 329 K. The star shows the average equilibrium structural relaxation time $\tau_{eq}$ obtained from the analysis of the aging experiment. For comparison, the structural relaxation time $\tau_{struct}$ from mechanical spectroscopy is plotted (pluses) [54]. The line is a fit of $\tau_{diel}(T)$ in equilibrium with a VFT law leading to $\tau_0 = 1.6 \times 10^{-15}$ s, B = 1720 K, and $T_{VF}$ = 276 K.

In order to demonstrate the spectral location of the aging data of xylitol and CKN, in Figs. 3 and 4 the broadband dielectric spectra of these materials are shown. In Fig. 3 loss spectra of xylitol are presented for various temperatures. In comparison to earlier works [13,22,23,24,25,26], the present data extend the investigated maximum frequencies by at least two decades. The spectra reveal the well-known strong shift of the α-relaxation peak with temperature and the β-relaxation, which is merged with the α-relaxation emerging at the lower temperatures. At the highest frequencies, despite the rather strong scatter of the data at low temperatures and the gap in the data between about 15 and 40 GHz, there are significant indications of a loss minimum, indicated by the dotted lines. It seems to shift towards lower frequencies with decreasing temperature. The closed symbols represent three spectra measured during the aging experiment performed at 243.4 K, again revealing a continuous decrease of the loss during aging [cf. Fig. 1(f)]. Due to the long measurement time needed for frequency sweeps extending to very low frequencies, the α-peak could not be covered in these spectra.

In Fig. 4, broadband modulus spectra for CKN are shown [8,28,29,30]. CKN is a prototypical ionic-melt glass former. In these materials, whose dielectric response is dominated by ionic conductivity, it is customary to plot the imaginary part of the complex dielectric modulus, $M^* = 1/\varepsilon^*$ ($\varepsilon^*$ the complex dielectric permittivity) [31]. The typical peaks shifting through the frequency window with temperature are observed and in addition at high frequencies, a minimum shows up. In Fig. 4, compared to the previously published spectra [8,28,29] additional data at 287 K and 350 K are provided. Furthermore, here we include two curves collected during the aging experiment, the one obtained 100 s after quickly cooling down to 329 K and the "fully aged" curve measured after keeping the sample at this temperature for more than ten weeks. It becomes obvious that during aging the M"-peak shifts to lower frequencies. Interestingly, for the fully aged curve at frequencies above about 3 kHz, M"($\nu$) exhibits a deviation from the power-law decrease at the high-frequency flank of the relaxation peak.

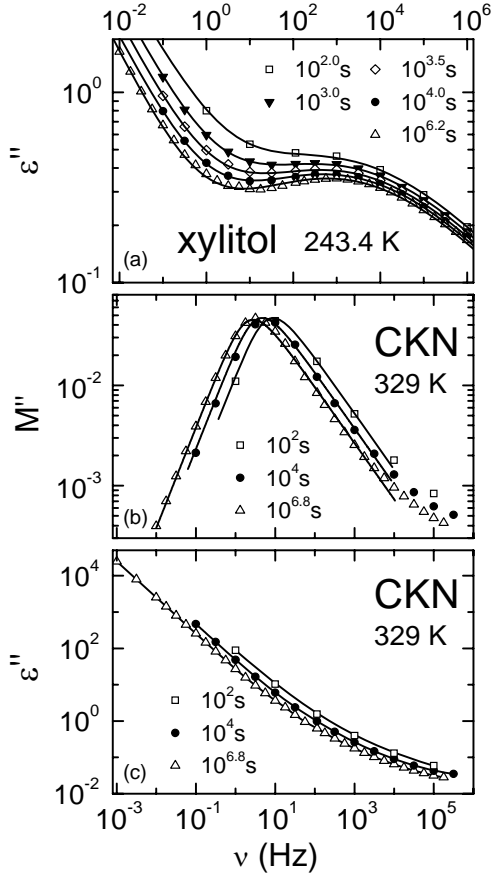

Fig. 5. Spectra of the dielectric loss of xylitol (a) measured at 243.4 K and of the modulus (b) and loss (c) of CKN measured at 329 K and various aging times. The lines in (a) are fits using the sum of a power law and the CC function. The lines in (b) are fits with the CD function. The lines in (c) are drawn to guide the eye. The experimental errors are of similar magnitude as the width of the symbols.

In Fig. 5 the variation of the spectra during aging are shown in more detail for xylitol (a) and CKN [(b) and (c)]. For xylitol the loss spectra and for CKN both, the imaginary part of the modulus [Fig. 5(b)] and the dielectric loss [Fig. 5(c)] are given. In xylitol immediately after reaching 243 K (uppermost curve), the loss shows an overall decrease with increasing frequency and a distinct shoulder with nearly constant $\varepsilon$" at about 100-1000 Hz. During aging, it develops into a well-pronounced peak. In CKN in M"($\nu$) a peak shows up, which continuously shifts to lower frequencies with increasing aging time. For all times an excess-wing like feature is seen. In contrast, in the loss a $1/\nu$ decrease (slope -1 in the double-logarithmic plot) up to about 1 kHz is detected, followed by a weaker decrease tending to saturate at the highest frequencies. The main effect of the aging seems to be a downward shift of the $\varepsilon$"($\nu$) curves.

## 4. Discussion

It seems a natural approach to fit time-dependent data of glass-forming materials as those shown in Fig. 1 with a stretched-exponential or Kohlrausch-Williams-Watts (KWW) law [32], given by $\exp[-(t/\tau)^\beta]$. The exponent $\beta$ (termed stretching or width parameter) takes into account the commonly found non-exponentiality of glass-forming matter, which can be ascribed to a distribution of relaxation times [33]. Adding some parameters to account for the saturation values at very short and long times, one arrives at [11,34]:

$$y(t_{age}) = (y_{st} - y_{eq}) \exp\left[-\left(t_{age}/\tau_{age}\right)^{\beta_{age}}\right] + y_{eq} \quad (1)$$

Here y denotes the measured quantity ($\varepsilon'$, $\varepsilon$", or M" in our case), $y_{st}$ and $y_{eq}$ are its values in the limit of zero and infinite times, respectively, and $t_{age}$, $\tau_{age}$, and $\beta_{age}$ are the aging time, aging relaxation-time, and aging width-parameter, respectively.

It is natural to use a KWW law because it is well established for the description of the structural relaxation in time-domain measurements, performed in thermodynamic equilibrium. In those experiments, the time-dependent reaction of the samples to an excitation is monitored, e.g. by measuring the stress relaxation after applying a constant strain to a rod made of the investigated material or the polarization after applying an electric field to a capacitor filled with the material. Also for experiments in frequency domain, the KWW function (i.e., its Fourier transform) is often used to fit the spectra. In the present aging experiments the excitation is given by the fast cooling step to a temperature below $T_g$. It is reasonable to assume that in this type of experiment the relaxation is determined by the same structural relaxation processes as in the equilibrium experiments. In both cases the molecules rearrange to accommodate for the external "force", i.e. the lower temperature in the aging case or the applied strain or electric field in the equilibrium cases. Thus one can expect that the values of $\tau_{age}$ and $\beta_{age}$ are identical to those of the corresponding quantities extracted from the higher-temperature equilibrium experiments if extrapolated to the lower temperatures of the aging experiments. However, as was shown by us for the ε"(t) data of Fig. 1 [15], when applying eq. (1), values of $\tau_{age}$ and $\beta_{age}$ result that markedly differ from the extrapolated equilibrium results. This also holds for the analyses of aging data using eq. (1) reported in [11,13,35]. Especially, the width parameter $\beta_{age}$ is found to be much smaller than $\beta$ obtained from equilibrium experiments. This leads to the unreasonable conclusion that the heterogeneity causing the distribution of relaxation times [33] is different for structural relaxation after a downward temperature step or after an external excitation in equilibrium.

The reason for this inconsistency is that eq. (1) contains a fundamental error: It assumes that the relaxation time is constant. However, during aging the structure of the glass rearranges (leading to a variation of the density) to finally arrive at the structure that is "correct" for the measuring temperature. As the relaxation time depends on the glass structure it also must vary during this structural rearrangement; in the present case of a downward temperature jump, leading to a density increasing with time, τ(t) also should increase. Such a concept was introduced already long ago, to account for the so called non-linearity effects arising in the response of glass-forming materials to temperature steps [36]. It led to the introduction of a fictive temperature, which characterizes the non-equilibrium structure of the aging glass. In the Tool-Narayanaswamy-Moynihan formalism [9,36,37] the aging-induced variation of physical quantities is traced back to the time dependence of the fictive temperature $T_f$, leading to a time-dependent $\tau_{age}$, and by introducing an additional non-linearity parameter. It is employed on a regular basis by glassmakers to predict the changes of glass properties after temperature jumps and during annealing. However, its application is not straightforward and involves some uncertainties, e.g., by requiring assumptions concerning the dependence of the relaxation time and the measured quantity on $T_f$.

The Tool-Narayanaswamy-Moynihan formalism also can describe experiments involving complex thermal histories. For the simple temperature-step experiments performed in the present work, we account for the time dependence of $\tau_{age}$ by using a simpler approach, recently proposed by us in [12]. It assumes that during aging after a downward temperature jump, the relaxation rate $v_{age} = 1/(2\pi\tau_{age})$ itself decreases with a KWW law, whose relaxation time is given by $\tau_{age}$ itself. The relaxation rate is a good measure of the α-peak frequency. In the modulus spectra of Fig. 5(b) is can be directly seen that the peak frequency reduces during aging and also for xylitol, if comparing the spectra from the aging experiment with the equlibrium ones (Fig. 3), such a scenario is the most likely one to explain the observed variation below about 1 Hz. Thus, analogous to eq. (1) we write:

$$v_{age}(t_{age}) = 1 / \left[2\pi \tau_{age}(t_{age})\right] \quad (2)$$
$$= (v_{st} - v_{eq}) \exp\left\{-\left[t_{age} / \tau_{age}(t_{age})\right]^{\beta_{age}}\right\} + v_{eq}$$

As was noted in [12,15] this ansatz can be easily solved by recursion. It implies that the stretching (quantified by $\beta_{age}$) remains unaffected by aging, i.e. that time-temperature superposition is valid. As the shift of $T_f$ during aging is only few K for these experiments, this assumption is justified. By putting the resulting time-dependent $\tau_{age}(t_{age})$ instead of a constant $\tau_{age}$ into eq. (1), good fits of the ε"(t) data of Fig. 1 and the M"(t) data of Fig. 3 are achieved (solid lines). In contrast to fits with a time-independent $\tau_{age}$ [11,13,15], with this approach the data can be described using the same width parameter $\beta_{age}$ as in equilibrium measurements [12,15,]. Moreover, the parameters $v_{st}$, $v_{eq}$, and $\beta_{age}$ are identical for all curves measured at different frequencies and the equilibrium relaxation times $\tau_{eq} = 1/(2\pi v_{eq})$, deduced from the aging experiments, well match the extrapolated structural relaxation times measured in equilibrium above $T_g$. The additional ε'(t) data shown in Figs. 1(a), (c), and (e) also were fitted in this way. We used the same parameters $\beta_{age}$ and $\tau_{age}(t)$ as obtained from the fits of the loss data [12]. Very good fits of the experimental data could be reached in this way, especially if having in mind that there are only two free parameters per curve, namely ε'$_{st}$ and ε'$_{eq}$, the limiting values for zero and infinite times.

In Fig. 1 the loss data of xylitol [Fig. 1(f)] stand out by showing a crossing of the curves for different frequencies. Three spectra obtained during aging, plotted together with the broadband spectra in Fig. 3 reveal the reason for this behavior. Obviously, the frequencies of the aging curves in Fig. 1(f) are located in the regime of different spectral processes. For example, at 0.1 Hz [circles in Fig. 1(f)] the data points were obtained at the high-frequency flank of the α-relaxation. In contrast, at 1 kHz (triangles down) and above (stars and triangles right), the behavior is dominated by the β-relaxation. Thus from Fig. 1(f) we conclude that the variation of ε"(t) in the β-relaxation regime is significantly weaker than in the α-relaxation regime. Obviously, the main effect of the aging on the α-relaxation is a shift of the peak to low frequencies. Based on literature data, for the β-relaxation it seems not so clear if the aging affects the β-relaxation time $\tau_\beta$ or reduces the peak amplitude (or both) [3,13,18,38,39]. In Fig. 5(a) the xylitol spectra at different aging times are fitted with the sum of a power law for the high-frequency flank of the α-peak and a phenomenological Cole-Cole (CC) function [40] for the β-peak. In agreement with the findings in [13], we are able to fit these data using a fixed β-relaxation time and a β-relaxation strength decreasing with time [lines in Fig. 5(a)]. However, the data can also be fitted with $\tau_\beta$ to some extent increasing or decreasing during aging and a significant statement on its time dependence based on the present data is not possible. But irrespective of the question how the *β*-relaxation is affected by the aging, the fact that in Fig. 1(e) and (f) the curves for high and low frequencies can be well described by the same aging parameters clearly demonstrates that the time dependence of this variation is determined by the structural α-relaxation. During aging the structural "environment" felt by the relaxing entities varies with the time constant $\tau_{age}$. This variation of the structural environment influences the β-relaxation, either by shifting its peak or by reducing its amplitude, however in both cases with a relaxation time given by $\tau_{age}$.

The dashed lines in Fig. 3 are the results of fits using the sum of a Cole-Davidson (CD) [41] or a Havriliak-Negami (HN) [42] function to account for the α-process and a CC function for the β-process. The resulting average relaxation-times [43] are provided in the inset (open circles). Similar analyses were performed, e.g., in [22,24], our results extending the available temperature range. In addition, we include $\tau_{eq}$ obtained from the aging experiment (closed circle). According to the scenario developed above, this aging relaxation time should be identical to the structural α-relaxation time. Thus it can be used to significantly extend the $\tau_\alpha(T)$ curve (at least if decoupling effects are neglected) and indeed it matches nicely the results at higher temperatures. The line shows a parameterization of $\tau_\alpha(T)$ obtained in that way by a Vogel-Fulcher-Tammann (VFT) law $\tau_\alpha = \tau_0 \exp[B/(T-T_{VF})]$ [1] with $\tau_0 = 6.4 \times 10^{-14}$ s, $B = 1290$ K, and $T_{VF} = 210$ K.

Concerning the β-relaxation times shown in the inset of Fig. 3, first it should be noted that at high temperatures the accuracy of $\tau_\beta(T)$ suffers from the strong superposition of α- and β-peak and the results at T > $T_g$ are of reduced significance only. Also, one should mention that it is not so clear if a simple additive superposition of different contributions to $\varepsilon''(\nu)$ is really justified, especially if there is a strong frequency overlap [44]. Neglecting for the moment these problems, sufficiently below $T_g$ $\tau_\beta(T)$ of xylitol follows an Arrhenius law, as is documented for many type B glass formers. However, close to $T_g$ ($1000/T_g \approx 4.1$ K$^{-1}$) it seems to become nearly constant and even to exhibit a minimum. Taking into account some recent reports [24,38,45,46], one may speculate that this could be a general behavior of type B glass formers. It can be described by the so-called "minimal model" (MM) by Dyre and Olsen [47]. Above $T_g$, $\tau_\beta(T)$ of xylitol starts to exhibit a much stronger temperature dependence and rather closely follows $\tau_\alpha(T)$ with a $\tau_\alpha/\tau_\beta$ ratio of about 30 at 260 K and 3 at room temperature (inset of Fig. 3). While the uncertainty of the data points in the region above $T_g$ is rather high, their deviation from Arrhenius behavior is significant. Such a deviation from Arrhenius behavior of $\tau_\beta(T)$ agrees with the findings in typical type A systems [48,49] were the β relaxation does not show up as a peak or shoulder but leads to an excess wing only [6]. It can be understood in the framework of the coupling model, assuming that its primitive relaxation time is located close to the JG β-relaxation time as was reported for numerous glass formers (e.g., [18,26,39,49]).

The $\tau_\beta(T)$ curve in the inset of Fig. 3 also may give a clue about the only weak or even absent time dependence of $\tau_\beta$ during aging [Fig. 5(a)]: With T = 243.4 K (1000/T ≈ 4.1 K$^{-1}$), the aging experiment was performed just in the region of only weak temperature dependence of $\tau_\beta(T)$ and thus it seems reasonable that $\tau_\beta(t)$ remains constant when the fictive temperature decreases during aging. But of course this requires that this weak temperature dependence really is an intrinsic equilibrium property. In addition, one may also speculate that the unexpected decrease of $\tau_\beta$ during aging at 230 K reported in [13] could be connected with the minimum in $\tau_\beta(T)$. However, according to our data, 230 K is already in the region of conventional temperature dependence of $\tau_\beta$, but one should have in mind that these data points were obtained far away from equilibrium and may strongly depend on thermal history.

Before leaving the xylitol spectra of Fig. 3, we want to briefly remark that the existence of a loss minimum at very high frequencies, as indicated by the dotted lines, is a well-established fact in a variety of glass formers [2,7,8,20,48,50]. There the minimum was found to be too shallow to be explained by a simple superposition of α-peak/excess wing and the microscopic excitations showing up in the infrared region, which clearly points to contributions from a fast process in this region. For a variety of materials it was shown [2,7,8,20,48,51] that in many aspects this minimum follows the predictions of mode coupling theory [52]. But also alternative explanations were proposed, e.g., in terms of the extended coupling model [53] assuming a nearly constant loss contribution. The present spectra do not provide sufficient data points to enable a quantitative evaluation of the minimum in terms of model predictions and current work is in progress to fill the frequency gap at 15 - 40 GHz and extend the spectra to higher frequencies.

Concerning the presented broadband spectra in CKN (Fig. 4), detailed analyses of the α-relaxation and fast dynamics were already reported by us in some earlier works [8,28,29,48]. The inclusion of the two spectra measured during aging at 329 K in Fig. 4 facilitates the rationalization of the mentioned unusual behavior of the 1 Hz and 10 Hz curves in Fig. 2(b). As revealed by Figs. 4 and 5(b), at the sub-$T_g$ temperature of 329 K still peaks in $M''(\nu)$ are observed in the available frequency window. At first glance, this is at odds with expectation because at the glass temperature the relaxation time should be of the order of 100 - 1000 s and thus the peak frequency located below the lowest measured frequency, just as in the case of xylitol (Fig. 3). However, it is well documented that in CKN the conductivity relaxation times determined from $M''(\nu)$ decouple from the structural relaxation times at low temperatures (see, e.g., [30]). Thus in this special case it is also possible to collect aging data at the low-frequency flank of the α-peak. As the peak shifts to lower frequencies during aging [Fig. 5(b)], $M''(t)$ in this region increases with time, as observed for the 1 Hz curve in Fig. 2(b). As shown in Fig. 2(b), where the lines represent fits by equations (1) and (2), also this curve can be fitted. This is achieved by using $y_{eq} > y_{st}$ in eq. (1). Figs. 4 and 5(b) also reveal that at 10 Hz a transition from the peak to the high-frequency flank of the α-peak takes place during aging, which in Fig. 2(b) leads to the indication of a maximum in $M''(t_{age})$ at short times. As discussed in [12,15], if during aging such a transition between different spectral regimes occurs, the simple description with eqs. (1) and (2) must fail.

In contrast to $M''(\nu)$, in $\varepsilon''(\nu)$ of CKN [Fig. 5(c)], no peaks show up, due to the strong contributions from charge transport, the dc conductivity leading to a $1/\nu$ dependence of $\varepsilon''(\nu)$ for $\nu < 10$ Hz (see also our broadband loss spectra of CKN in [28,29,48]). Thus the time-dependent ε'' shown in Fig. 2(a) decreases continuously. The lines in Fig. 2(a) are fits using the same parameters of eq. (2) as for the modulus, being identical for all frequencies. Also here a good match of the experimental data could be achieved, again demonstrating the rather universal applicability of our approach. In contrast to the $M''(t)$ curve at 10 Hz [Figs. 2(b)], in $\varepsilon''(t)$ at this frequency there is no transition between two spectral regions during aging and the behavior always is dominated by the dc conductivity [Fig. 5(c)]. Therefore in the ε'' representation also the 10 Hz curve can be well described by eqs. (1) and (2). The good fits of $\varepsilon''(t)$ for all frequencies show that, as long as the measured quantity is in some way influenced by the structural environment and there is no transition between different regions, its time dependence can be described by eq. (1) using the time-dependent $\tau_{age}$ of eq. (2). As mentioned, in the present case the loss reflects the ionic charge transport in CKN. The dynamics of the mobile ions certainly is faster than the structural relaxation time, leading to the well-known decoupling effects. However, it is reasonable that the time-dependent variation of this dynamics during aging is determined by the structural relaxation time, i.e. by the much slower dynamics of the immobile ions forming the structure.

When considering the picture developed above, dielectric aging experiments should provide the possibility to measure the structural relaxation time even in systems where conventional dielectric spectroscopy suffers from decoupling effects. In the inset of Fig. 4, the temperature dependences of the dielectric relaxation times $\tau_{diel}$ and the structural ones ($\tau_{struct}$), obtained from mechanical measurements by Howell *et al.* [54], are given. Both can be parameterized by a VFT law (shown as line for $\tau_{diel}$), except for the much weaker temperature dependence of $\tau_{diel}$ at low temperatures occurring in non-equilibrium below $T_g$ [30]. The triangles depict the relaxation times deduced from the fits of the two aging curves shown as solid symbols in Fig. 4, the upper triangle representing the result from the fully aged (i.e., equilibrium) curve. This value satisfactorily matches the points at higher temperatures. From the same aging experiment, when analyzing the time dependent data, also the structural equilibrium relaxation time, namely $\tau_{eq}$, was deduced and revealed to be more than six decades larger than $\tau_{diel}$. It is indicated by the star in the inset of Fig. 4 and, as expected, provides a reasonable extension of the $\tau_{struct}(T)$ curve [54].

Finally, we briefly want to point out that the fully aged M"-spectrum at 329 K, which was measured with higher precision than our earlier spectra in CKN [28,29,30], shows evidence for an excess-wing like feature at frequencies above about 3 kHz. The presence of an excess wing also in ionic-melt glass formers was already suspected in [28], based on the older CKN data and especially on M"(ν) measured in $[Mg(NO_3)_2]_{0.44}[KNO_3]_{0.56}$ (MKN). In this context it is interesting that recently in so-called ionic liquids evidence for secondary relaxations was obtained from dielectric experiments [55]. It can be speculated that the excess wing in CKN and MKN also is caused by a secondary relaxation, nearly submerged under the α-relaxation peak. Such a scenario is well established for type A glass formers as, e.g., glycerol or PC [6,18,22,49,56].

## 5. Conclusion

In summary, in the present paper we have provided detailed information on the aging-time dependence of different dielectric quantities as dielectric constant, loss, and modulus in a variety of glass formers. The comparison with the broadband spectra shown for xylitol and CKN reveals that, depending on measuring frequency, the aging was performed at different spectral regions, namely α-process and β-process for xylitol and the left and right wing of the α-peak for CKN. The experimental data and analysis provided in the present paper corroborate the notion promoted in [12,15] that, irrespective of the dynamic process prevailing in the investigated frequency region, the aging dynamics is always determined by the structural relaxation process. This agrees with the conclusion from a recently developed free-energy landscape model that in the reequilibration of glass formers during aging there is no other time scale than that of the structural α-relaxation [57]. We also have demonstrated that different dielectric quantities show the same time dependence during aging, determined by the structural relaxation time and the same heterogeneity as in equilibrium.

The aging experiments provided in xylitol, together with the high-frequency equilibrium measurements allows for a considerable extension of the known temperature dependence of the relaxation time in this glass former. In CKN there is the interesting situation that by the dielectric aging experiment, both the dielectric relaxation time (from the analysis of the fully aged spectrum) and the structural relaxation time (from the analysis of the aging-time dependent data), both differing by more than six decades, could be determined.

**Acknowledgements**

We thank C.A. Angell, R. Böhmer, R.V. Chamberlin, J.C. Dyre, G.P. Johari, N.B. Olsen, and R. Richert for illuminating discussions. We thank U. Schneider for performing part of the measurements.